\documentclass[aps,prl,twocolumn,superscriptaddress,showpacs,preprintnumbers,amsmath,amssymb,nofootinbib,longbibliography]{revtex4-1}
\usepackage[
colorlinks=true, 
pdfstartview=FitV, 
linkcolor=blue, 
citecolor=magenta, 
urlcolor=blue, 
bookmarks=true,
bookmarksnumbered=true,
pdftitle={},
pdfauthor={}t
]{hyperref}

\usepackage{amssymb,amsmath,array,verbatim,ascmac,amsthm,bm}
\usepackage[dvipdfmx]{graphicx}
\usepackage{color}

\newcommand\sect[1]{{\it #1.}---}

\newcommand{\beq}{\begin{eqnarray}}
\newcommand{\eeq}{\end{eqnarray}}
\newcommand{\p}{\partial}

\newcommand{\bpm}{\begin{pmatrix}}
\newcommand{\epm}{\end{pmatrix}}

\newcommand{\ba}{\left(\begin{array}}
\newcommand{\ea}{\end{array} \right)}

\newcommand{\rmi}{\mathrm{i}}
\newcommand{\diff}{\mathrm{d}}

\begin{document}

\title{Instantons in Chiral Magnets}

\author{Masaru Hongo}
\email{masaru.hongo(at)riken.jp}
\address{Department of Physics, 
and Research and 
Education Center for Natural Sciences, 
Keio University, 4-1-1 Hiyoshi, Yokohama, Kanagawa 223-8521, Japan
}
\address{iTHEMS, RIKEN, Wako, Saitama 351-0198, Japan}

\author{Toshiaki Fujimori}
\email{toshiaki.fujimori018(at)gmail.com}
\address{Department of Physics, 
and Research and 
Education Center for Natural Sciences, 
Keio University, 4-1-1 Hiyoshi, Yokohama, Kanagawa 223-8521, Japan
}

\author{\\ Tatsuhiro Misumi}
\email{misumi(at)phys.akita-u.ac.jp}
\address{Department of Mathematical Science, Akita 
University, 1-1 Tegata-Gakuen-machi,
Akita 010-8502, Japan
}
\address{Department of Physics, 
and Research and 
Education Center for Natural Sciences, 
Keio University, 4-1-1 Hiyoshi, Yokohama, Kanagawa 223-8521, Japan
}
\address{iTHEMS, RIKEN, Wako, Saitama 351-0198, Japan}

\author{Muneto Nitta}
\email{nitta(at)phys-h.keio.ac.jp}
\address{Department of Physics, 
and Research and 
Education Center for Natural Sciences, 
Keio University, 4-1-1 Hiyoshi, Yokohama, Kanagawa 223-8521, Japan
}

\author{Norisuke Sakai}
\email{norisuke.sakai(at)gmail.com}
\address{Department of Physics, 
and Research and  
Education Center for Natural Sciences, 
Keio University, 4-1-1 Hiyoshi, Yokohama, Kanagawa 223-8521, Japan
}
\address{iTHEMS, RIKEN, Wako, Saitama 351-0198, Japan}

\begin{abstract}

We exhaustively construct instanton solutions and 
elucidate their properties in one-dimensional 
anti-ferromagnetic chiral magnets
based on the $O(3)$ nonlinear sigma model description of spin chains 
with the Dzyaloshinskii-Moriya (DM) interaction.
By introducing an easy-axis potential and a staggered magnetic field, 
we obtain a phase diagram consisting of ground-state phases with
two points (or one point)  
in the easy-axis dominant cases, 
a helical modulation at a fixed latitude of the sphere,
and a tricritical point allowing helical modulations 
at an arbitrary latitude.
We find that instantons (or skyrmions in two-dimensional 
Euclidean space) appear as composite solitons 
in different fashions in these phases: temporal domain walls 
or wall-antiwall pairs (bions) 
in the easy-axis dominant cases, 
dislocations (or phase slips) with fractional instanton numbers 
in the helical state, 
and isolated instantons and calorons
living on the top of the helical modulation at the tricritical point.
We also show that the models with DM interaction 
and an easy-plane potential 
can be mapped into those without them, providing 
a useful tool to investigate the model 
with the DM interaction.
\end{abstract}

\maketitle
\sect{Introduction} 
Topological excitations (topological solitons and instantons) 
play key roles in various systems from 
particle physics~\cite{Rajaraman:1982is,Coleman,Manton:2004tk} 
and cosmology~\cite{Vilenkin:2000jqa}
to condensed matter systems~\cite{Volovik2003universe,Nelson2002}.
Amongst various examples, 
magnets allow
magnetic skyrmions and domain walls in spin systems~\cite{Bogdanov1989,Bogdanov1994,Rossler2006,Binz}, 
and in particular,
chiral magnets with the Dzyaloshinskii-Moriya (DM) interaction~\cite{Dzyaloshinsky1958,Moriya:1960zz} 
is a representative where topological excitations play 
a pivotal role for applications to nano-devices such as magnetic memories. 
Recent theoretical and experimental developments have confirmed that 
there exists the so-called chiral soliton lattice phase 
--- aligning helical domain walls ---in one-dimensional spin chains~\cite{Togawa2012,Kishine2015,Togawa2016}.
More recently, a skyrmion lattice in two-dimensional (2D) chiral magnets 
and associated peculiar transport phenomena have been 
experimentally observed~\cite{Muhlbauer2009,Yu2010,Heinze2011,Nagaosa2013}.  
Furthermore, magnetic monopoles have been recently paid much attentions.
They appear for unwinding a skyrmion line in a skyrmion lattice 
\cite{Milde1076} and form a stable crystal in certain parameter region \cite{Kanazawa2016}.
One of recent interesting theoretical developments 
might be the finding of a critical coupling at which the strength of DM interaction and 
Zeeman magnetic field or magnetic anisotropy are balanced 
\cite{Barton-Singer:2018dlh,Adam:2019yst,Adam:2019hef,Schroers:2019hhe}, 
analogous to superconductors at the critical coupling between types I and II.
In this case, it allows so-called Bogomol'yi-Prasad-Sommerfield (BPS) topological solitons
\cite{Bogomolny:1975de,Prasad:1975kr}, {\it i.\,e.}
the most stable configurations 
with a fixed boundary condition 
(or a topological sector), 
which were originally found for magnetic monopoles 
and other topological solitons in high energy physics 
and now are realized in superconductors at the critical coupling.
 
Despite of such experimental and theoretical developments, 
instantons \cite{Belavin:1975fg,Polyakov:1975yp} 
(see Refs.~\cite{Rajaraman:1982is,Coleman,Manton:2004tk})
-- classical solutions of Euclidian field theory 
and one of the most crucial theoretical concepts 
to understand physical properties of quantum systems --have never been studied 
thus far in chiral magnets.
They represent nonperturbative quantum effects 
coming from nontrivial saddle point solutions of the Euclidian path integral.
With the help of instantons, we can understand several key results of 
physical systems such as 
a ground-state property of non-abelian gauge theory~\cite{tHooft:1976snw,tHooft:1976rip,Jackiw:1976pf,Callan:1976je,Schafer:1996wv} 
and nonlinear sigma models.

In this Letter, we work out, for the first time, instantons in chiral magnets.
After presenting a phase diagram, 
we exhaustively provide instanton solutions 
in one-dimensional anti-ferromagnetic spin chains 
with the DM interaction: 
temporal domain walls, 
domain wall-antidomain wall pair (called bions),
vortices or dislocations 
as fractional instantons (called merons),
and BPS instantons and calorons at the critical coupling.
Although most experiments focus on ferromagnetic chiral magnets, 
anti-ferromagnetic chiral magnets also exhibit rich behaviors 
as we show below.

\sect{Model and ground state}
Magnetic spin texture in quantum spin-chain is described 
in the continuum limit by a unit spin vector $n^a\, (a=1,2,3)$ 
with $\sum_a n^a n^a=1$. 
The energy functional for one-dimensional spin-chain 
involving the DM interaction with the strength $\kappa$ 
in addition to the kinetic term, easy-axis potential and staggered 
magnetic field is given at low-energy as a form of the $O(3)$ 
or ${\mathbb C}P^1$ sigma model: 
\begin{align}
 E[n] =
 \int \diff x \bigg[
 &\frac{(\partial_x n^a)^2}{2}  
 +\kappa (n^1\partial_x n^2-n^2\partial_x n^1) 
 \nonumber \\
 &+ \mu \frac{1-(n^3)^2}{2} + B n^3\bigg].
\label{eq:energy-functional}
\end{align}
For $\mu>0$, the potential favors for $n^a$ to point to 
the north ($n^3=1$) or the south ($n^3=-1$) pole (easy-axis). 
For $\mu<0$, it favors the equator ($n^3=0$) (easy-plane). 
The term $Bn^3$ is a staggered magnetic field, 
but,  for simplicity, we call $B$ as a magnetic field.

\begin{figure}[t]
 \centering
 \includegraphics[width=0.8\linewidth]{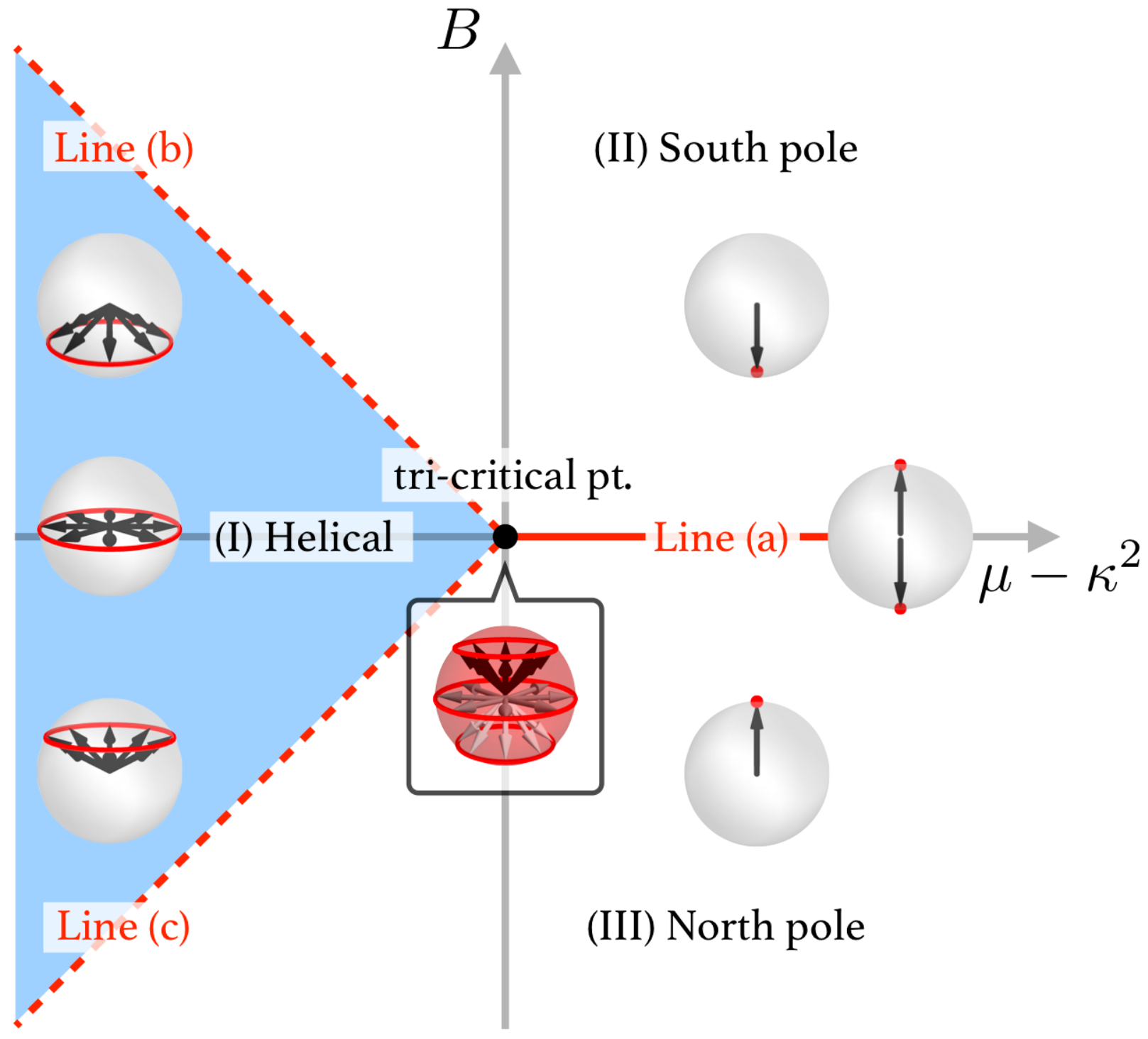}
 \caption{
 Phase diagram of the $O(3)$ nonlinear sigma model with 
 the DM interaction in the meand-field level.}
\label{fig:phase}
\end{figure}
\begin{table}[t]
  \begin{tabular}{c|c|cccc} 
      & OPM $G/H$ & $\pi_0$ & $\pi_1$ & $\pi_2$ & $\pi_3$ \\ \hline
  (I) & $S^1$ & 0 & $\mathbb{Z}$ & 0 & 0\\
  (a) & $\mathbb{Z}_2$ & $\mathbb{Z}_2$ & 0 & 0 & 0\\
  Tri-critical pt. & $O(3)/O(2) \simeq S^2$ & 0 & 0 & $\mathbb{Z}$ & $\mathbb{Z}$ \\
    (II) (III) (b)(c)& 1pt & 0 & 0 & 0  &0\\
 \end{tabular}  
 \caption{
The OPMs and associated homotopy groups.
}
\label{tab:symmetry} 
\end{table}

Since the DM interaction can be regarded 
as a background gauge field~\cite{Schroers:2019hhe}, 
the energy can be rewritten 
by defining a covariant derivative 
$D_x n^a \equiv \partial_x n^a - \kappa \epsilon^{3ab} n^b$ as 
\beq
\!\!\!
E[n] = \! \int \! \diff x \!
\left[ \frac{(D_x n^a)^2}{2} 
+ (\mu -\kappa^2) \frac{1-(n^3)^2}{2}+Bn^3\right] \! . 
\label{eq:Bogomolnyi-bound}
\eeq
Since the first nonnegative term must vanish for a ground state, 
we obtain 
\beq
D_x n^a=0 \quad (a=1,2,3), 
\label{eq:Bogomolnyi-eq}
\eeq
which has the helical-state solution given by
\beq
n^1+ \rmi n^2 = Ae^{- \rmi \kappa x}, \; n^3= \pm \sqrt{1-|A|^2} 
\label{eq:helical-ground-state}
\eeq
with $A\in {\mathbb C}$ satisfying $|A|\le 1$. 
This is the spatially modulated state, 
where $n^a$ rotates at a constant latitude of the $S^2$ target space 
with the wave number $\kappa$ along the spatial direction $x$. 
The actual minimum of $E[n]$ depends on the value of the 
easy-axis potential $\mu$ and the magnetic field $B$. 
By examining the minimum of the potential as a function of $n^3$, 
we find three critical lines emanating from the tri-critical point 
at $\mu=\kappa^2,\, B=0$ as illustrated in Fig.~\ref{fig:phase}: 
the line (a) along $B=0,\, \mu >\kappa^2$, 
the line (b) along $\kappa^2-\mu=B>0$, and 
the line (c) along $\kappa^2-\mu=-B>0$. 
The ground state of the chiral magnet is helical states 
in the region (I) the south pole in the region (II) 
and the north pole in the region (III), 
respectively (See Fig.~\ref{fig:phase}).  
At the tri-critical point $B=0,\,\mu=\kappa^2$, 
all the helical states with $|A| \le 1$ become the ground states with the same energy. 
The topology of the order parameter manifold (OPM) 
is different in various regions 
as summrized in Table \ref{tab:symmetry}: 
$S^2$ at the tri-critical point, 
two discrete points (the north and south poles) along the line (a), 
$S^1$ in the region (I), 
a point (north pole) in the region (II) including the line (b), 
and a point (south pole) in the region (III) including the line (c). 
It is interesting to observe that the DM interaction tends to 
favor easy-plane configuration, so that helical ground states 
can be ground states even in the presence of the easy-axis potential with $\mu>0$. 

The order of phase transition for the critical lines are as follows: 
The first-order phase transition occurs on the line (a), 
since the global minimum of the energy jumps 
from one local minimum to the other across (a). 
The second-order phase transition occurs on the lines (b) and (c), 
where the second derivative of the ground state energy density 
with respect to $B$ is discontinuous.
It is also notable that the tri-critical point, 
which is a switching point of the first- and second-order transitions, 
has a larger symmetry as we will discuss in details later.

In order to explore instanton solutions 
for the anti-ferromagnetic material\footnote{
Effective theory of ferromagnetic material involves the first 
order term in time derivative, instead of the second order. 
}, 
we introduce the imaginary time $\tau$ 
and consider the following Euclidean 2D Lagrangian $(i=0,1)$
\begin{equation}
 \begin{split}
  \mathcal{L}
  =& 
\frac{1}{2}(\partial_i n^a)^2
 + \kappa (n^1\partial_x n^2 - n^2\partial_x n^1)
  \\
& +\frac{\mu}{2} [1- (n^3)^2 ] +B n^3.
\label{eq:euclid_action_DM}
 \end{split}
\end{equation}
Note that this Euclidean 2D model should also be useful to describe 
the energy density of a 2+1D magnetic material 
(both ferromagnetic and anti-ferromagnetic) which 
has an anisotropic uniaxial DM interaction 
only in one spatial direction $x$ rather than two spatial directions. 
It is also convenient to use the stereographic projection of 
the target space $n^a\in S^2$ to a complex plane $v\in {\mathbb C}$ through 
\beq
v=\frac{n^1+ \rmi n^2}{1+n^3}, \quad n^3=\frac{1-|v|^2}{1+|v|^2}.
\label{eq:stereographic}
\eeq
The $O(3)$ sigma model with the DM interaction becomes 
\begin{align}
 \mathcal{L}
 =&
 \frac{2\{|\partial_i v|^2
 +\rmi \kappa v \overset{\leftrightarrow}{\p} \bar v +\mu|v|^2\}}{(1+|v|^2)^2}
+ B\frac{1-|v|^2}{1+|v|^2}
\label{eq:lagrangian_complex}
\end{align}
with $v \overset{\leftrightarrow}{\p} \bar v = v \partial_x \bar v-\bar v\partial_x v$.
The instanton number density $\rho_Q$ 
is defined by 
\beq 
\rho_Q = 
\frac{1}{4\pi}
\epsilon_{abc} n^a \partial_x n^b \partial_\tau n^c 
= \frac{1}{\pi}
\frac{\partial_z v \partial_{\bar z} \bar v
-\partial_{\bar z} v \partial_{ z} \bar v}{(1+|v|^2)^2},
\label{eq:instanton-number_complex}
\eeq
where we used the complex coordinates $z=x+ \rmi \tau,\, \bar z=x- \rmi \tau$.
The intergration of this quantity in the Euclidean 2D 
space 
yields the instanton number charactrizing 
the second homotopy group $\pi_2$. 
Below, we show that instanton solutions exist in all the phases, 
sometimes as composite objects 
 even when the OPS does not have a nontrivial $\pi_2$.

\sect{Domain walls on the line (a)}
In this case 
the OPM consists of two points,
suggesting that there exists a domain wall solution 
connecting these two discrete ground states. 
For instance, we can impose a boundary condition 
$n_3(\tau, x=\pm\infty)=\pm 1$
at the left and right spatial infinities. 
For a $\tau$-independent configuration, 
we can make a Bogomol'nyi completion~\cite{Bogomolny:1975de,Prasad:1975kr} 
of the energy to find 
\begin{align}
\!\!\! \int \diff x \, {\cal L}
 = 2 \int \diff x \bigg[
\frac{|(\partial_x + \eta ) v |^2}{(1+|v|^2)^2} + \partial_x \frac{\sqrt{\mu-\kappa^2}}{1+|v|^2} \bigg]
\label{eq:BPS_wall_x}
\end{align}
with $\eta = \sqrt{\mu - \kappa^2} + \rmi \kappa$. 
Since the first term is positive semi-definite, 
the surface term provides the lower bound of the energy:
\beq
\int \diff x \, {\cal L}
\ge 2\sqrt{\mu-\kappa^2}
\left[\frac{1}{1+|v|^2} \right]_{x=-\infty}^{x=\infty}. 
\label{eq:bound_wall_x}
\eeq
The equality holds when the BPS equation
\beq
(\partial_x+ \rmi \kappa) v+\sqrt{\mu-\kappa^2} \, v=0
\label{eq:BPSeq_wall_x}
\eeq
is satisfied.
This equation gives the following domain wall solution 
with a complex integration constant $C$ 
\beq
v=C e^{- \rmi \kappa x-\sqrt{\mu-\kappa^2}x}, \quad C \in {\mathbb C}.
\label{eq:wall_sol_x}
\eeq
This is the lowest energy configuration 
satisfying the boundary condition.
This does not carry an instanton number
in contrast to the temporal case discussed below.

\begin{figure}[t]
 \centering
 \includegraphics[page=1, width=0.99\linewidth]{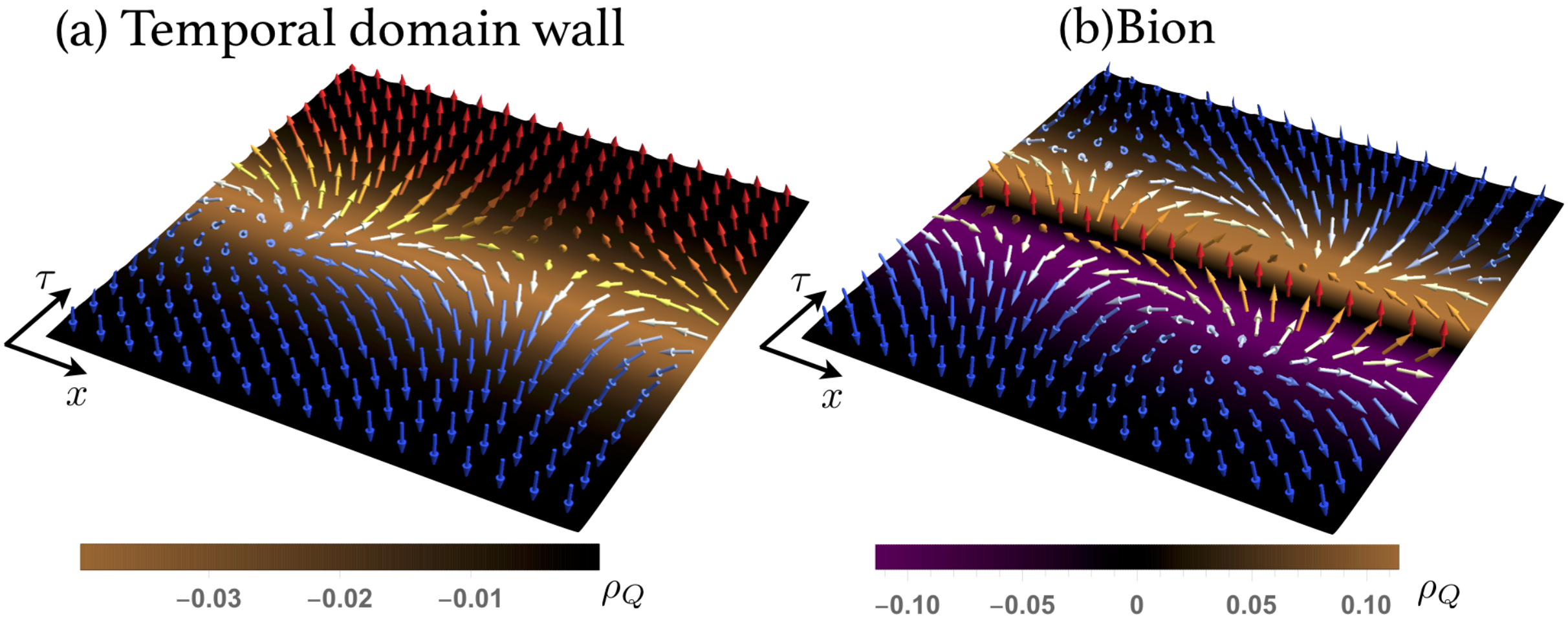}
 \caption{
 (a) A temporal domain wall configuration of $\bm{n}$ 
 interpolating two degenerate ground states, as an instanton solution. 
 (b) Domain wall-antidomain wall pair (bion). 
 }
\label{fig:DW}
\end{figure}

Similarly we can construct a domain wall solution 
in the temporal direction, regarded as an instanton, 
for the boundary condition $n_3(\tau=\pm\infty, x)=\pm1$. 
The Bogomol'nyi bound for the action per unit $x$ is obtained 
by replacing $x \to \tau$ in Eq.~(\ref{eq:bound_wall_x}). 
This bound is saturated if the BPS equations 
\beq
(\partial_x+ \rmi \kappa) v=0, \quad 
\partial_\tau v+\sqrt{\mu-\kappa^2} \, v=0
\label{eq:BPSeq_wall_x}
\eeq
are satisfied.
The domain wall solution is 
\beq
v=Ce^ {-\rmi\kappa x-\sqrt{\mu-\kappa^2}\tau},  \quad
C \in {\mathbb C}
\label{eq:wall_sol_x}
\eeq
with unit instanton number 
and action $2\sqrt{\mu-\kappa^2}$ 
per length $\Delta x = 2\pi /\kappa$. 
It is interesting to note  
that as shown in Fig.~\ref{fig:DW} (a), helical states show up 
only in the vicinity of the wall
even though it is hidden in the ground states%
 \footnote{  
Since 
in 2+1D, a domain wall carrying a skyrmion number 
(equivalent to the instanton number
for Euclidean 2D) density 
induced by a modulating $U(1)$ phase on the wall 
is called a domain wall skyrmion \cite{Nitta:2012xq,Kobayashi:2013ju}, 
we may call this solution a domain wall instanton.
} ($n^3=\pm 1$).

\sect{Bions in the phases (II) and (III)}
If the magnetic field is turned on ($B\not=0$)
in the region (II) or (III),  
we find that the linear term in $n^3$ allows 
non-BPS wall-antiwall solutions, 
so-called bion solutions~\cite{Fujimori:2016ljw,Fujimori:2017oab}. 
For example, the simplest solution 
with a single wall-antiwall pair is given by
\beq
v = e^{-\rmi\kappa(x-x_0)} \frac{\sqrt{B}}{\omega} \sinh \omega (\tau-\tau_0), 
\eeq
where $\omega = \sqrt{\mu-\kappa^2+B}$ and 
$\tau_0$ and $x_0$ are the position moduli parameters. 
These solutions play a vital role in nonperturbative effects 
and resurgence theory 
\cite{Dunne:2012zk,Dunne:2016nmc,Misumi:2014jua,Misumi:2014bsa,Misumi:2015dua,Fujimori:2016ljw,Fujimori:2017oab,Fujimori:2017osz} 
recently being extensively considered in field theory.

\begin{figure}[htb]
 \centering
 \includegraphics[page=1, width=0.99\linewidth]{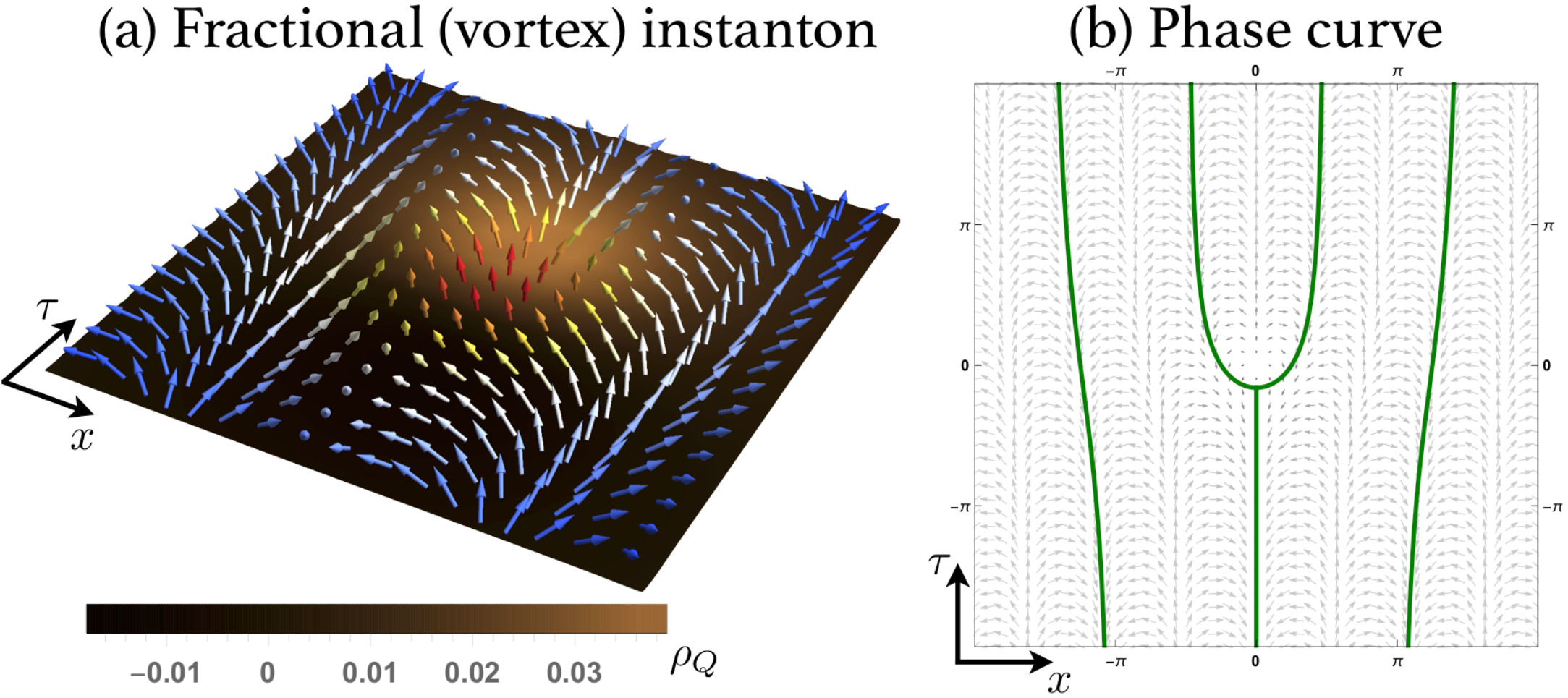}
 \caption{(a) Fractional instanton (meron) ($\mu=3,~\kappa=2,~B=-1/2$).
 (b) Fractional instanton as a dislocation or phase slip. 
 Solid lines indicate the contours with $\arg v = - \pi/2$ and arrows denote $(n_1,n_2)$.}
\label{fig:fractional_instanton}
\end{figure}

\sect{Fractional instantons in helical state (I)}
In this case, the OPM is $S^1$ at latitude 
$\sqrt{(\kappa^2-\mu+B)/(\kappa^2-\mu-B)}$. 
Therefore, we expect to find a vortex 
characterized by a non-trivial winding number around $S^1$. 
It carries a fractional instanton number \cite{Nitta:2011um} 
as is known for skyrmions with easy-plane potential 
\cite{Jaykka:2010bq,Kobayashi:2013aja},
and is sometimes called a meron.
A single-winding configuration takes the form
\beq
v = e^{- \rmi \kappa x} \frac{z}{|z|} f(|z|),
\eeq
where $z=x+\rmi\tau, \bar z=x-\rmi\tau$ are the complex coordinates 
and $f(|z|)$ is a profile function 
satisfying $f(0) = 0$ and $f(\infty) = \sqrt{(\kappa^2-\mu+B)/(\kappa^2-\mu-B)}$. 
Numerically solving field equations we find a fractional instanton. 
Fig.\,\ref{fig:fractional_instanton} (a) shows the profile of 
a fractional instanton, 
which also exhibits a dislocation of the phase 
(or a phase slip) as 
shown in Fig.\,\ref{fig:fractional_instanton} (b). 

\sect{Instantons in the tri-critical point}
The richest array of instantons is obtained at the tri-critical 
point $\mu=\kappa^2, \,B=0$, where the Euclidean 2D action 
is bounded by the instanton number $Q = \int d^2 x \, \rho_Q$ as 
\beq
\int \diff^2 x \, {\cal L}_{\mathrm{cr}} \ge 
\pm (4\pi Q+\int \diff^2 x \, \kappa\partial_\tau n^3). 
\label{eq:Bogomolnyi-bound}
\eeq
The bound is saturated if and only if the BPS equation
\beq
\partial_\tau n_a\mp \epsilon_{abc} n^b D_x n^c = 0 
\quad \mbox{ or } \quad
\partial_{\bar z} v = - \rmi \kappa v/2 
\label{eq:BPS-eq}
\eeq
is 
satisfied. Let us consider solutions with
a nonnegative instanton number $Q$ (by taking the upper sign of the bound).
Hence all the BPS solutions are obtained as 
\beq
v(z,\bar z) = e^{-\rmi\kappa x} w(z), 
\label{eq:BPS-sol-complex}
\eeq
with an arbitrary holomorphic function $w(z)$. 

\begin{figure}[b]
 \centering
 \includegraphics[width=01.0\linewidth]{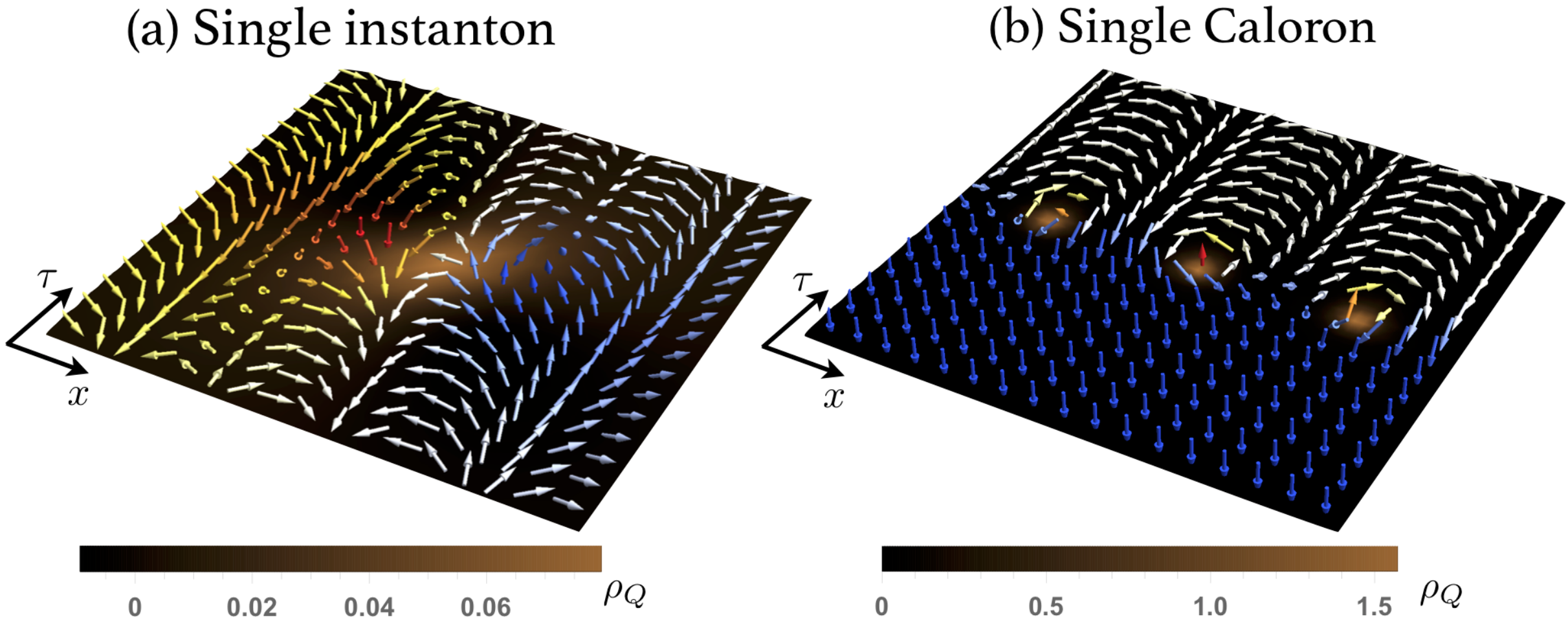}
 \caption{
 Configurations of $\bm{n}$ for (a) single BPS instanton solution:
 $p(z)/q(z) = (z-z_0)/(z+z_0)$, and (b) single BPS caloron solution:
 $p(e^{\rmi\frac{2\pi z}{L}})/q(e^{\rmi\frac{2\pi z}{L}})
 = \lambda (e^{\rmi\frac{2\pi(z-z_0)}{L}}-1)$.
 }
\label{fig:Critical-BPS}
\end{figure}

The BPS equation has the general $l$-instanton BPS solution given 
by a rational function of degree $l$
\beq
v=e^{-\rmi \kappa x} \frac{p(z)}{q(z)}, 
\label{eq:BPS_l-instanton}
\eeq
where $p(z),\, q(z)$ are mutually coprime polynomials of degree 
$m,\, n$ with $l=\max(m,n)$ [Fig.~\ref{fig:Critical-BPS} (a)]. 
We can also construct general $l$-caloron (periodic instantons)-solutions 
in the periodicity interval $x \sim x+L$ as
\beq
v=e^{-\rmi \kappa x} \frac{p(e^{\rmi\frac{2\pi z}{L}})}{q(e^{\rmi\frac{2\pi z}{L}})} 
\label{eq:BPS_l-caloron}
\eeq
with a rational map of degree $l$ [Fig.~\ref{fig:Critical-BPS} (b)].
Let us note that all these caloron solutions satisfy a twisted 
boundary condition 
$v(x+L, \tau)=e^{-\rmi\kappa L} v(x, \tau)$. 
By changing a moduli parameter, 
this caloron can be split into a pair of a temporal domain wall and antidomain wall with fractional instanton numbers, separated at arbitrary distance with the same energy  
\cite{Eto:2004rz, Eto:2006mz, Eto:2006pg}.  

\sect{Equivalence theorem}
Let us now show that 
there exists a one-to-one mapping between 
the above model \eqref{eq:euclid_action_DM}
and the usual $O(3)$ sigma model.
For that purpose, 
inspired by the helical ground states \eqref{eq:helical-ground-state}, 
we define new variables%
\footnote{
When our system is put in the finite interval $0 \leq x \leq L$, 
we also need to take account of the change of the boundary condition.
}
$\hat n^a$ as 
\beq
n^1+ \rmi n^2= (\hat n^1+ \rmi \hat n^2)e^{- \rmi \kappa x}, \quad 
n^3=\hat n^3.
\label{eq:hat_n}
\eeq
In terms of the new variables, the $O(3)$ sigma model with 
the DM interaction can be rewritten 
into that without the DM interaction 
${\cal L}_{\mathrm{woDM}}(\hat n) = {\cal L}
(n)$ :
\begin{eqnarray}
{\cal L}_{\mathrm{woDM}}(\hat n)
 &=&
\frac{1}{2}(\partial_\mu \hat n^a)^2+\frac{\mu-\kappa^2}{2} [1-(\hat n^3)^2] +B \hat n^3. \quad\quad
\label{eq:pure_CP1}
\end{eqnarray}
Note that the strength of the easy-axis potential is reduced from the original one. 
From this equivalence theorem, we find that all instanton solutions 
in the $O(3)$ sigma model with the DM interaction has one-to-one 
correspondence with those of the $O(3)$ sigma model without the DM interaction. 
This indicates that the original model \eqref{eq:euclid_action_DM} 
possesses a kind of hidden symmetry 
(known as modified symmetry~\cite{Ohashi:2017vcy,Takahashi:2017ruq}), 
which enables us to compactly 
summarize our finding on the ground states and instanton 
solutions (See Table~\ref{tab:symmetry}).
Nevertheless, it should be emphasized that the physical variable $n^a$ 
must be used to find the real magnetic texture of chiral magnets.

\sect{Summary and Discussion}
We have clarified possible instanton solutions for the one-dimensional 
anti-ferromagnetic spin chain in the presence of the DM interaction. 
Depending on the phases, 
we have exhausted all possible instanton solutions including
temporal domain walls with the instanton number
density distributed 
in the temporal direction, 
vortices (dislocations or phase slips) as fractional instantons and
BPS instantons and caloron at the critical coupling.  
We have also shown that the model with the DM interaction is equivalent to the model without the DM interaction.

Our results have ienergy. 
mplications both for theoretical and experimental researches.
The instanton solutions in chiral magnets, which were not discussed before, give a novel theoretical insight into the anti-ferromagnetic spin chains and our methodology to obtain them based on the equivalence theorem can be applied broadly in the related studies.
The phase diagram in Fig.~\ref{fig:phase} with the variety of instantons can help us to understand physics which would be observed in future experiments on chiral magnets with controllable DM interaction \cite{Siegfried:2015,Koretsune:2015,Belabbes:2015,Ma:2016} or easy-axis potential.

There are several interesting avenues related to this work.
While we have considered one-dimensional chiral magnets in this letter, 
we can generalize our approach to higher-dimensional systems~\cite{Barton-Singer:2018dlh,Adam:2019yst,Adam:2019hef,Schroers:2019hhe}.
In particular, the 2D 
model allows a topologically conserved 
skyrmion current and Hopf terms, thus involves rich theoretical structures.
Although we have only focused on the ground state at the mean field level and instanton solutions 
interpolating them, it is also interesting to consider generic quantum aspects of 
the systems; e.\,g. a generalization of the Haldane 
conjecture~\cite{Haldane:1983ru,Haldane:1982rj}
in chiral anti-ferromagnetic chains, and deconfined quantum criticality 
in $2+1$-dimensional systems~\cite{Haldane:1988zz,Read:1989zz,Senthil2004}.
The possible 't Hooft anomaly --- field theoretical manifestation of 
the Lieb-Schultz-Mattis theorem~\cite{Lieb:1961fr,Affleck:1986pq} --- together with 
semi-classics analyses including resurgence theory in the sigma models~\cite{Dunne:2012zk,Dunne:2016nmc,Misumi:2014jua,Misumi:2014bsa,Misumi:2015dua,Fujimori:2016ljw,Fujimori:2017oab,Fujimori:2017osz}
will shed light on the detailed quantum aspects of chiral magnets.

\begin{acknowledgments}
 This work is supported by MEXT-Supported Program for the Strategic Research Foundation
 at Private Universities ``Topological Science'' (Grant No. S1511006) 
 and by
 the Japan Society for the 
 Promotion of Science (JSPS) Grant-in-Aid for Scientific Research
 (KAKENHI) Grant Number  18H01217.  
 The authors are also supported in part by 
 JSPS KAKENHI Grant Numbers 
 18K03627 (T.\ F.), 19K03817 (T.\ M.), and 16H03984 (M.\ N.).
 M.H. is also partially supported by RIKEN iTHEMS Program (in particular, iTHEMS STAMP working group).
 The work of M.N. is also supported in part by a Grant-in-Aid for Scientific Research on Innovative Areas ``Topological Materials Science'' (KAKENHI Grant No. 15H05855) from MEXT.
\end{acknowledgments}

\bibliography{helical_DM.bib}

\end{document}